\documentstyle[12pt]{article}
\newcommand{\be}{\begin{equation}}
\newcommand{\nn}{\nonumber}
\newcommand{\bea}{\begin{eqnarray}} 
\newcommand{\eea}{\end{eqnarray}}
\newcommand{\ba}{\begin{array}}
\newcommand{\ea}{\end{array}}
\newcommand{\ee}{\end{equation}}

\expandafter\ifx\csname mathbbm\endcsname\relax

\else

\fi \textheight 22cm \textwidth 15cm \topmargin 1mm \oddsidemargin
5mm \evensidemargin 5mm

\newcommand{\beas}{\begin{eqnarray*}}
\newcommand{\eeas}{\end{eqnarray*}}
\newcommand{\bes}{\begin{equation*}}
\newcommand{\ees}{\end{equation*}}

\newcommand{\lf}{\left}
\newcommand{\ri}{\right}
\newcommand{\f}{\frac}

\def\tr           {\mbox{\rm tr}\,}

\def\i2           {\mbox{$\frac{i}{2}$}}

\def\lt   {{\tilde l}}

\def\tF           {\tilde F}

\def\del           {\delta}

\def\ep           {\epsilon}

\def\la           {\lambda}

\def\om           {\omega}

\def\rh           {\rho}

\def\pl           {\partial}

\def\ran          {\rangle}
\def\lan          {\langle}

\begin{document}

\begin{titlepage}
\hfill \vbox{
    \halign{#\hfil         \cr
           } 
      }  
\vspace*{20mm}
\begin{center}
{\LARGE \bf{{Correction to baryon spectrum in holographic QCD}}}\\ 

\vspace*{15mm} \vspace*{1mm} {Ali Imaanpur}

\vspace*{1cm}

{\it Department of Physics, School of Sciences\\ 
Tarbiat Modares University, P.O. Box 14155-4838, Tehran, Iran\\
Email: aimaanpu@modares.ac.ir}\\
\vspace*{1mm}

\vspace*{1cm}

\end{center}

\begin{abstract}
We study solitons in the context of Sakai-Sugimoto-Witten holographic model of QCD. In the large 't Hooft coupling limit, flat space instantons give an approximate description of baryons. We provide an ansatz through deforming the BPST instanton by zero modes and obtain a subleading correction to the effective action. We also consider the quantum mechanical description of the collective coordinates. The correction term appears as a perturbation to the potential of the Hamiltonian and hence we can derive a first order correction to the mass spectrum of baryons.    
\end{abstract}

\end{titlepage}

\section{Introduction}
Since its proposal, there has been a great deal of effort to extend AdS/CTF duality \cite{MAL} to some more realistic theories such as QCD. Using the duality, one can use the weakly coupled supergravity to learn about the behaviour of the gauge theory at the strong coupling limit. Therefore, in this way, one hopes to better understand some strong coupling phenomena in QCD such as confinement and chiral symmetry breaking \cite{KAR, ERD, MAT}. 

One of such holographic models of QCD is the Sakai-Sugimoto-Witten model which has been successful in many respects \cite{WIT98T, SUG}, see \cite{ERD2} for a review. For instance, it has modeled the chiral symmetry breaking in QCD and has approximately reproduced the spectrum of hadrons based on the Witten's idea of identifying baryons with the solitons/instantons of the corresponding model \cite{WIT79}. In this model one considers a stack of D8-$\overline{{\rm D8}}$ probe branes propagating in the background of $N_c$ D4-branes. Upon compactification over an $S^4$ and setting the corresponding components of the gauge fields to zero one is left with an effective 5-dimensional action. Solitons of this effective theory are to be identified with the baryons in QCD. 

In fact, the static solitons are the classical solutions of the 4-dimensional Euclidean reduced theory. In \cite{HATA, YI}, it has been argued that, in a particular limit, the BPST instanton sit at the minimum of the action. In this paper, we follow this idea and present an ansatz that has a lower action. A small deformation of the instanton along its zero modes produces another solution which has the same action up to quadratic order. At quartic order, the action deviates from that of instanton by a positive-definite term. We exploit this observation to construct our ansatz, for which, as we will see, the effective action attains a lower value. This happens because the effective action is not $SO(4)$ invariant; the symmetry along the $z$ direction is broken by the warp factors. In the next step, we consider quantum mechanics of the collective coordinates and take the new term in the action as a small perturbation to the potential. In so doing, we calculate the corrections to the energy eigenvalues that have already been evaluated in \cite{HATA}. We compare the results with the experimental data and comment on the importance of these corrections.

\section{Effective Action}
There are several approaches to discuss solitons in the Sakai-Sugimoto-Witten model \cite{POMA, SEK, CHER, COLAN, RAAM, PRE}.  Here we follow one of the earliest ones developed in \cite{HATA}. We use the results derived in there and refer to this paper for more detail. 
In this approach, one starts considering time independent field configurations in 5 dimensions. A coordinate and field rescaling by $\la$, the 't Hooft coupling, allows one to  consistently expand the field equations in powers of $1/\la$ for large $\la$. The leading field equation is just the 4-dimensional pure Yang-Mills equation which has instantons as its exact solutions. The next order equations can conveniently be described by an effective action which in addition to $SU(N)$ gauge fields involves the time component of a $U(1)$ gauge field $\hat{A}_0$. 
Let us then write the whole effective action up to order one. The Yang-Mills part is
\be
S_{\rm{YM}}=k\int d^3x dz\, \tr\! \lf(\f{1}{2} h(z) F_{ij}F^{ij} + k(z) F_{kz}F^{kz}\ri) -\f{k}{2\la}\int d^3x dz\, (\pl_\mu \hat{A}_0)^2\, ,   \label{FF}
\ee
where 
\be
k=a\la N_c =\f{\la N_c}{216\pi^3}\, , 
\ee
with $a=1/216 \pi^3$. We have also used $\mu =(i,z)$, with $i,j,\ldots =1,2,3$, as a collective 4-dimensional index. The warp factors $h(z)$ and $k(z)$ are defined as follows
\be
h(z)=(1+z^2/\la)^{-1/3}\approx 1-z^2/3\la \, , \ \ \ \ \ k(z)=1+z^2/\la\, .
\ee
Note that the subleading terms in $1/\la$ explicitly break the translational invariance along the $z$ direction. There is also a Chern-Simons term in the effective action in this order
\be
S_{\rm{CS}}=-\f{N_c}{16\pi^2}\, \ep_{ijk}  \int d^3x dz \, \hat{ A_0}\,  \tr (F_{ij}F_{kz})\, . \label{CS}
\ee

Let us start our discussion with instanton solutions in Yang-Mills theory. They are self-dual solutions having the absolute minimum 
value of the action. The gauge potential of the BPST solution, in the regular gauge, is given by
\be
A_i=\f{1}{\xi^2+\rh^2}\lf((z-Z)\tau_i -\ep_{ijk}(x_j-X_j)\tau_k \ri) \, ,\ \ \ \ \  A_z=-\f{(x_i-X_i)\tau_i}{\xi^2+\rh^2}\, ,\label{BPST}
\ee
with the field strength\footnote{As for the conventions, we use  $A_\mu=A_\mu^a \tau^a$, $D_\mu A_\nu =\pl_\mu A_\nu + i[A_\mu , A_\nu]$, and $F_{\mu\nu}=\pl_\mu A_\nu - \pl_\nu A_\mu + i[A_\mu , A_\nu]$, where $\tau^a$ are the Pauli matrices. }
\be
F_{ij}=\f{2\rho^2}{(\xi^2+\rh^2)^2}\ep_{ijk}\tau_k\, , \ \ \ \ \ \ F_{zk}= \f{2\rho^2}{(\xi^2+\rh^2)^2}\tau_k\, ,
\ee
where $Z$ and $X^i$ indicate the position and $\rho$ the size of the instanton. And
\be
\xi^2 = (z-Z)^2 + r^2 +\rh^2\, ,
\ee
with $r^2=\sum_{i=1}^3 (x_i-X_i)^2$. 
The ansatz we make is through using the zero modes. Since the action has broken the translation invariance along the $z$ direction, it will be useful to look at a zero mode due to the variation along the $Z$ direction of the instanton location. Accompanied with a gauge transformation to keep the gauge condition $D_\mu \del A^\mu = 0$, the zero mode reads:
\be
\del A_i = \f{B}{(\xi^2 +\rho^2)^2} \tau_i\, , \ \ \ \ \del A_z =0\, , \label{ANS}
\ee
for $B$ a constant to be determined by minimizing the action. Being a zero mode, $\del A_i$ satisfies the following equations:
\bea
&& D^2\del A_i -2i [F_{\mu i} , \del A^\mu] =0\, ,\nn \\
&& D^2 \del A_z -2i [F_{iz} , \del A_i] = 0\, .\label{ZER}
\eea
The above equations can be derived by varying the action around an instanton up to quadratic order. First we note
\be
\tilde{F}_{\mu\nu}=F_{\mu\nu}+\del F_{\mu\nu} = F_{\mu\nu}+D_\mu\del A_\nu -D_\nu\del A_\mu +i[\del A_\mu , \del A_\nu]\, .
\ee
Hence, we have
\bea
S \!\! &=&\!\!  \int d^4x \, \tr\! (\tilde{F}_{\mu\nu} \tilde{F}^{\mu\nu}) =  \int d^4x \, \tr\!  \lf ({F}_{\mu\nu} {F}^{\mu\nu} + 4 F_{\mu\nu} D^\mu\del A^\nu\ri)\, \nn\\
\!\! &+&\!\! 2\int d^4x \, \tr\!  \lf( D_\mu\del A_\nu (D^\mu\del A^\nu -D^\nu\del A^\mu ) + iF_{\mu\nu}[\del A^\mu , \del A^\nu] \ri)\nn \\
\!\! &+&\!\! \int d^4x \, \tr\!  \lf(4i D_\mu\del A_\nu  [\del A^\mu , \del A^\nu]  - [\del A_\mu , \del A_\nu][\del A^\mu , \del A^\nu]\ri)\, .\label{TIL}
\eea
 The linear term in $\del A_\mu$ vanishes since we are expanding around a solution. By part integration, the quadratic terms read
\bea
&& 2\int d^4x \, \tr\!  (  \del A^\nu (-D^2 \del A_\nu +D_\mu D_\nu\del A^\mu) + iF_{\mu\nu}[\del A^\mu , \del A^\nu] )\nn \\
&&= 2\int d^4x \, \tr\!  (  \del A^\nu (-D^2 \del A_{ \nu} + 2i[F_{\mu\nu},\del A^\mu ]  ))\, ,\label{QUAD}
\eea
where in the second equality we have used the gauge condition $D_\mu \del A^\mu=0$, and
\be
[D_\mu , D_\nu] \del A_\rh = i[F_{\mu\nu} , \del A_\rh]\, .
\ee
The operator appearing in (\ref{QUAD}) is the zero mode operator that we started from in (\ref{ZER}). 

In expanding around the instanton in (\ref{TIL}), in order to get a function of $B$ we have kept all the terms as the linear and quadratic terms vanish for the zero modes. Now, taking $A^\mu=A^\mu_{\rm {BPST}}+\del A^\mu$, with $A^\mu_{\rm {BPST}}$ and $\del A^\mu$ given by (\ref{BPST}) and (\ref{ANS}), respectively, and plugging it in (\ref{TIL}) we arrive at
\bea
\f{1}{192}\, \tr\! \lf(\f{1}{2} \tF_{ij}\tF^{ij} + \tF_{kz}\tF^{kz}\ri)\!\! &=&\!\! \f{ \rh^4}{4(\xi^2 +\rh^2)^4} \nn \\
\!\! &-&\!\! \f{ B \rh^2 (z-Z) }{(\xi^2 +\rh^2)^5} + \f{  B^2 ((z-Z)^2 - \f{\rh^2}{4} +\f{r^2}{6})}{(\xi^2 +\rh^2)^6} \nn \\
\!\! &+&\!\!  \f{ B^3 (z-Z)}{(\xi^2 +\rh^2)^7}  +\f{ B^4}{8(\xi^2 +\rh^2)^8} \, . \label{FIJ}
\eea  
Note that the asymmetry between $z$ and $x_i$ coordinates in the above expression reflects the broken $SO(4)$ symmetry in our ansatz (\ref{ANS}). This will be important when we compute the second part of the action which includes $z^2$ factors. Integration of (\ref{FIJ}) results in
\be
\int d^4x\,  \tr\!\lf(\f{1}{2} \tF_{ij}\tF^{ij} + \tF_{kz}\tF^{kz} \ri)= 8\pi^2 +\f{4\pi^2 B^4}{7 \rh^{12}}\, .\label{AC1}
\ee
The absence of $B^2$ term in this expression explicitly shows that we are expanding around a zero mode. If we ignore the second integral in (\ref{FF}) for a moment, using (\ref{FIJ}), for the first part of the action we obtain
\bea
S_{\rm YM}\!\! &=&\!\! k \int d^4 x \, \tr\! \lf(\f{1}{2}(1-\f{z^2}{3\la}) \tF_{ij}\tF^{ij} + (1+\f{z^2}{\la})\tF_{kz}\tF^{kz}\ri) \nn \\
\!\! &=&\!\! 8\pi^2 k +\f{4\pi^2 k B^4}{7 \rh^{12}}+\f{8 \pi^2 k}{\la}\lf( \f{\rh^2}{6}+\f{Z^2}{3} -\f{B Z}{3\rh^2} +\f{B^2 (8 Z^2+3\rh^2)}{40 \rh^6} \ri) .\label{AC2}
\eea
Note that $B$ appears in the leading order of the action and therefore it is not a moduli. If the subleading terms were absent the minimum of the action would obviously occur at $B=0$. 

Let us then minimize action (\ref{AC2}) with respect to $B$. First, note that terms proportional to $B^2$ are of order ${1}/{\la}$, so they contribute to the subleading corrections upon minimization. Keeping the order one $B^4$ term and the order ${1}/{\la}$ term proportional to $B$ and minimizing with respect to $B$ results in
\be
B= \lf(\f{7Z \rh^{10}}{6\la}\ri)^{1/3} . \label{B}
\ee
Plugging this back into action (\ref{AC2}), we get
\be
S_{\rm YM}=8\pi^2 k + \f{8 \pi^2 k}{\la}\lf( \f{\rh^2}{6} + \f{Z^2}{3} \ri) -2\pi^2 k \lf(\f{7}{6}\ri)^{1/3} \lf(\f{Z \rh }{\la} \ri)^{4/3} .\label{ACN2}
\ee
The correction is of order $1/\la^{4/3}$, and this is bigger than the correction of $1/\la^2$ that would come from the expansion of $h(z)$ in (\ref{FF});
\be
h(z) =(1+z^2/\la)^{-1/3} \approx 1-z^2/3\la +2 z^4/9\la^2 + \cdots
\ee
so we only keep ${\cal O}(\la ^{-1})$ in the above expansion. Had we chosen another zero mode corresponding to a translation along $X^i$ (or a large gauge transformation) in our ansatz, we would only get positive-definite quadratic terms of order $1/\la$ with no linear terms in (\ref{AC2}). Therefore, minimization would give the trivial result of $B=0$. This is because action (\ref{FF}) has broken the symmetry only along the $z$ direction. 

The upshot of the discussion in this section is if we use the following ansatz
\be
A^i = A^i_{ \rm{BPST}} + \f{B}{(\xi^2 +\rho^2)^2} \tau^i\ , \ \ \ \ A^z=A^{z}_{\rm {BPST}}\, ,\label{SATZ}
\ee
with $B= \lf(\f{7Z \rh^{10}}{6\la}\ri)^{1/3}$, then, as is evident from (\ref{ACN2}), the effective action is always less than or equal to the effective action of a BPST instanton.

\section{Chern-Simons Contribution}
In this section we discuss the contribution of the Chern-Simons term to the action. From (\ref{FF}) and (\ref{CS}), the equation of motion for $\hat {A}_0$ becomes
\be
\pl^2 \hat {A}_0 = \f{1}{16 \pi^2 a}\, \ep_{ijk}\, \tr (F_{ij} F_{kz})\, .\label{A0}
\ee
With ansatz (\ref{SATZ}), for the RHS we obtain
\bea
\f{1}{192}\, \ep_{ijk}\,  \tr (F_{ij} F_{kz}) \!\! &=&\!\! \f{ \rh^4}{4 (\xi^2 +\rh^2)^4} -  \f{  B \rh^2 (z-Z) }{(\xi^2 +\rh^2)^5} \nn \\
\!\! &+&\!\!  \f{ B^2 ((z-Z)^2 - \f{\rh^2}{4} +\f{r^2}{6})}{(\xi^2 +\rh^2)^6} + \f{ B^3 (z-Z)}{(\xi^2 +\rh^2)^7}   \, .
\eea  
Using the Green's function of $\pl^2$ operator in (\ref{A0}), i.e.,
\be
G(x-y)=\f{1}{4\pi^2 (x-y)^2}\, ,
\ee
we can derive $\hat {A}_0$:
\bea
\hat {A}_0\!\! &=&\!\! \f{1}{8 \pi^2 a} \, \f{2 \rh^2 +\xi^2}{ (\xi^2 +\rh^2)^2} - \f{B}{8 \pi^2 a}\,  \f{(z-Z)(3\rh^2 +\xi^2)}{\rh^2 (\xi^2 +\rh^2)^3} 
\nn \\
\!\!&-&\!\! \f{B^2}{96 \pi^2 a\, \rh^4 (\xi^2 +\rh^2)^4}\lf[ 9\rh^4+\xi^4-4(z-Z)^2 \xi^2+ 4\rh^2(\xi^2-4(z-Z)^2)\ri] \nn \\
\!\!&+&\!\! \f{B^3 (z-Z)}{160 \pi^2 a\, \rh^8 (\xi^2 +\rh^2)^5}\lf[ 11\rh^6 -3\rh^2 \xi^4 +(\rh^2+\xi^2)^2(9\rh^2+2\xi^2\ri]  .\label{A1}
\eea

Having obtained $\hat{A}_0$, we can compute its contribution to the action. Upon integration, however, there will be no contribution linear in $B$. With no linear $B$ term in the Chern-Simons action, minimization with respect to $B$ gives the same result  we obtained in (\ref{B}). As $B$ is of order $\la^{-1/3}$, higher powers of $B$ in (\ref{A1}) will give corrections which are smaller than the subleading terms, and so we neglect them. Therefore, including the contribution from the first term in (\ref{A1}), for the complete action we obtain 
\be
S=8\pi^2 k + \f{8 \pi^2 k}{\la}\lf( \f{\rh^2}{6} + \f{1}{320\pi^4 a^2}\f{1}{ \rh^2}+ \f{Z^2}{3}\ri) -2\pi^2 k \lf(\f{7}{6}\ri)^{1/3} \lf(\f{\rho Z}{\la} \ri)^{4/3} .\label{COMPLETE}
\ee
In the next section we consider $\rho$ and $Z$ as approximate collective coordinates which are time dependent. The dynamics will be described by a quantum mechanical system which descends directly from the original 5-dimensional theory \cite{HATA}.   

\section{Quantization and Correction to Energy Eigenvalues}
In this section we use the results of \cite{HATA} referring to this paper for more detail. Treating $\rho$ and $Z$ as approximate collective coordinates, the second term of action (\ref{COMPLETE}) proportional to $1/\la$ will be the potential of our quantum mechanical problem. The kinetic terms are descending directly from the 5-dimensional action and the metric is induced from that of the moduli space of BPST instanton. The energy eigenvalues are given by \cite{HATA}: 
\be
E_{n_\rho, n_z, l}=8\pi^2 k +\sqrt{\f{(l+1)^2}{6}+\f{2}{15} N_c^2} \, + \f{2(n_\rho +n_z)+2}{\sqrt{6}}\, ,\label{EIG}
\ee
where $n_\rho, l$, and $n_z$ are integer quantum numbers associated with separable energy eigenfunctions of $\rho$ (together with global gauge transformations),  and $Z$, respectively. 

The third term of (\ref{COMPLETE}) is the subleading term in the large $\la$ limit, so we take it as a small perturbation to the potential of collective coordinates. Hence, we can work out the first order correction to the energy eigenvalues
\be
\Delta  E_{n_\rho, n_z, l} = -2\pi^2 a N_c \lf(\f{7}{6 \la}\ri)^{1/3} \lf\lan\rho^{4/3}  Z^{4/3} \ri\ran_{n_\rho, n_z, l}\, .\label{DELTA}
\ee
To compute these corrections, however, we will need the explicit energy eigenfunctions. The radial part of the wave function is given by
\be
R(\rho)=e^{-\f{m_y \om_\rh}{2}\rh^2}\, \rh^\lt \, F(-n_\rh, \lt+2; m_y \om_\rh \rh^2) \, ,
\ee
this together with the spherical harmonics on $S^3$, which we do not need their explicit expressions, constitute the wave function associated with $\rho$ and the global gauge transformations. In the above formula, $F(a,b;x)$ is the confluent hypergeometric function, whereas $\lt$ and $m_y \om_\rh$ are given by
\be
\lt \equiv -1 +\sqrt{(l+1)^2 +\f{4 N_c^2}{5}} \, ,\ \ \  m_y \om_\rh =\f{2}{9\sqrt{6}\pi} \, .
\ee
 The wave function of $Z$ part is just that of a simple harmonic oscillator
\be
\psi_{n_z}(Z) =\f{1}{\sqrt{2^{n_z} n_z!}} \, e^{-\f{m_z \om_z}{2} Z^2}\, H_{n_z} (\sqrt{m_z \om_z} Z)  \, ,
\ee
with $H_n(z)$ the Hermite functions, and $m_z \om_z =2/9\sqrt{6}\pi $. 

To proceed further, we also need to fix $M_{\rm{KK}}$ and $\la$ to compute the first few energy eigenvalues, including the subleading corrections. $M_{\rm{KK}}$ is the Kaluza-Klein mass scale in the theory which was set to one from the outset. To get it back all energy and mass formula need to be multiplied by this mass scale. Comparison with the meson data shows that $M_{\rm{KK}}=949\, $MeV and $\la\approx 16.6$ \cite{SUG}. However,  to have a best fit with the data from Particle Data Group (PDG) baryon summary in Tabel 1 \cite{PDG}, we have taken $\la=12$ and $M_{\rm{KK}}=500$ MeV. The results are brought in Tabel 2, where we have set $N_c=3$ in (\ref{EIG}) and (\ref{DELTA}). The two tables are as follows: 

\begin{table}[htbp!]	
\centering

\begin{tabular}{|l|c|c|c|c|c|c|c|}

\hline
$ (n_\rho , n_z)$ & (0,0) & (1,0)  & (0,1) & (1,1) & (2,0) , (0,2) &(2,1) , (0,3) & (1,2) , (3,0)  \\
 \hline
$N (l=1)$   & $940^+$     & $1440^+$      & $1535^-$     & $1655^- $    & $\!\! 1710^+\ , \ ? $ & $\!\! 2090^-_*$\  ,\ ? & $\!\! 2100^+_*\  ,\  ?$ \\
 \hline
${\sl \Delta} (l=3) $   & $1232^+$     & $1600^+$      & $1700^-$     & $1940^-_*$      & $\!\! 1920^+\  ,\  ?$ & ?\  ,\  ? & ?\  ,\  ?   \\
\hline

\end{tabular}

\label{tab1}

\caption{PDG Baryon Spectrum}
\end{table}

\begin{table}[htbp!]	
\centering

\begin{tabular}{|l|c|c|c|c|c|c|c|}

\hline
$ (n_\rho , n_z)$ & (0,0) & (1,0) , (0,1) & (1,1)\  ,\ (2,0) , (0,2) &(2,1) , (0,3) & (1,2) , (3,0)  \\
 \hline
$N (l=1)$   & $991^+$     & $1320^+ ,\  984^-$     & $1206^-  ,\ 1655^+ , \ 1116^+ $ & $1444^- ,\  1261^- $ & $1267^+ ,\  1995^+ $ \\
 \hline
${\sl \Delta} (l=3) $   & $1222^+$     & $1558^+ ,\ 1123^-$     & $1361^- ,\ 1898^+ ,\ 1193^+ $ & $1610^-  , \ 1280^- $ & $1365^+ , \ 2242^+$   \\
\hline

\end{tabular}

\label{tab2}

\caption{Baryon Spectrum with $\la=12$ and $M_{\rm{KK}}=500$ MeV}
\end{table}

\noindent
In the above tables $\pm$ superscripts refer to the parity of states, and the $*$ in Table 1 indicates that the evidence for the existence of that particular baryon is poor. The interpretation of the quantum numbers $(n_\rho , n_z)$ is proposed in \cite{HATA}. 

As we observe, in Table 2 the symmetry between $n_\rh$ and $n_z$ (seen in (\ref{EIG})) gets broken by $\Delta E$, and this is expected from Table 1.  Moreover, the contribution of $\Delta E$, being negative, is essential to get a better match with the experimental data. In particular, we get a good agreement with the experimental data whenever $n_z$ is zero. For non-zero values of $n_z$, however, the absolute value of the subleading corrections, $\Delta E$, become too large.
In fact, we get a better fit for non-zero $n_z$ if we choose $M_{\rm{KK}}=700\, $MeV. 

It is also worth mentioning that the leading contribution $E$ provides a good match with the data if one takes, as in \cite{HATA}, $M_0=8\pi^2 k=8\pi^2 a\la N_c \approx -0.3$, see Table 3. Unfortunately, this would correspond to an unrealistic value for $\la\approx -8.5$. Therefore, to make sense of the expansion around flat instantons and interpret them as baryons it is necessary to go beyond the leading order and take the contribution of $\Delta E$ into account.   

\begin{table}[htbp!]	
\centering

\begin{tabular}{|l|c|c|c|c|c|c|c|}

\hline
$ (n_\rho , n_z)$ & (0,0) & (1,0) , (0,1) & (1,1)\  ,\ (2,0) , (0,2) &(2,1) , (0,3) & (1,2) , (3,0)  \\
 \hline
$N (l=1)$   & $940^+$     & $1348^+ ,\  1348^-$     & $1756^-  ,\ 1756^+ , \ 1756^+ $ & $2164^- ,\  2164^- $ & $2164^+ ,\  2164^+ $ \\
 \hline
${\sl \Delta} (l=3) $   & $1240^+$     & $1648^+ ,\ 1648^-$     & $2056^- ,\ 2056^+ ,\ 2056^+ $ & $2464^-  , \ 2464^- $ & $2464^+ , \ 2464^+$   \\
\hline

\end{tabular}

\label{tab3}

\caption{Baryon Spectrum as computed in \cite{HATA} without correction, $\la=-8.5$ and $M_{\rm{KK}}=500$ MeV}
\end{table}


\end{document}